\documentclass[superscriptaddress,twocolumn,aps,prl,floatsfix]{revtex4}
\usepackage{graphicx}
\usepackage{amssymb,amsmath}
\usepackage{array}
\usepackage{dcolumn}

\begin{document}

\title{Large increase of the Curie temperature by orbital ordering control}

\author{Aymeric Sadoc, Bernard Mercey, Charles Simon, Dominique
  Grebille~\footnote{Deceased february 2009}, Wilfrid Prellier and
  Marie-Bernadette Lepetit}

\affiliation{CRISMAT, ENSICAEN-CNRS UMR6508, 6~bd. Mar\'echal Juin, 14050
   Caen, FRANCE}



 

\begin{abstract}
  Using first principle calculations we showed that the Curie temperature of
  manganites thin films can be increased by far more than an order of
  magnitude by applying appropriate strains.  Our main breakthrough is that
  the control of the orbital ordering responsible for the spectacular $T_C$
  increase cannot be imposed by the substrate only. Indeed, the strains, first
  applied by the substrate, need to be maintained over the growth direction by
  the alternation of the manganite layers with another appropriate
  material. Following these theoretical findings, we synthesized such
  super-lattices and verified our theoretical predictions.
\pacs{73.22.-f,73.21.Cd,75.70.-i}
\end{abstract}

\maketitle
Ferromagnetic super-lattices with very high $T_C$ present obvious interests
for electronic applications.  This is the first step required to design and
control magnetism in more elaborate structures.  Perovskite manganese oxides
of generic formula $\rm A_{1-x}B_xMnO_3$ (A being a trivalent cation and B a
divalent one) are metallic and ferromagnetic in the approximative range of
doping values $0.17<x<0.5$~\cite{LSMO-DPh}. These properties are directly
related to the $3d$ shell of the manganese atoms and can be easily explained
by the double-exchange mechanism. The Curie temperature is thus directly
related to the delocalization within the $e_g$ orbitals of the $\rm Mn^{3+}~/~
Mn^{4+}$ ions~\cite{GoodEnough}, that is to the manganese ions orbital
order. Finding a way to design such materials with the desired $3d$ orbital
order would thus open the way to the design of artificial materials with the
desired magnetic properties, and thus to novel magneto-electronic
applications, including spin-valve devices or non-volatile magnetic memory
working far above room temperature.

\begin{figure}[h]
\resizebox{6cm}{!}{\includegraphics{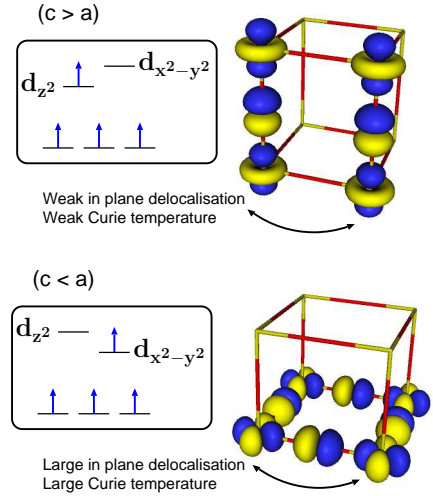}} \\
\caption{(Color online) Schematic representation of orbital ordering induced
  by cell distortions in a two dimensional system. ($c>a$) The elongation in
  the $\vec c$ direction favors the $d_{3z^2-r^2}$ orbital occupation and weak
  in-plane ferromagnetic coupling, the delocalisation being essentially in the
  $\vec c$ direction. ($c<a$) The contraction of the $c$ parameter favors the
  $d_{x^2-y^2}$ orbital occupation, strong in-plane ferromagnetic coupling and
  large Curie temperature.}
\label{fig:dist}
\end{figure}

The manganese $e_g$ shell is doubly degenerated (in a regular octahedron
environment) however only partly occupied. The $\rm Mn^{3+}$ ions are thus
Jahn-Teller active and small distortions of the oxygen octahedra can stabilize
one of the $e_g$ orbitals with respect to the other. If one can impose a
specific geometry around the Mn ions (for instance using synthesis in thin
films), it should thus be possible to choose between the $d_{3z^2-r^2}$ and
the $d_{x^2-y^2}$ orbitals which one will be occupied and which one will be
empty.  If the $d_{3z^2-r^2}$ is occupied, the double exchange will take place
essentially along the $\vec c$ direction, that is between $(\vec a,\vec b)$
planes (see figure~\ref{fig:dist}). On the contrary if the $d_{x^2-y^2}$ is
stabilized, the intra-layer double exchange will be very strong and the
inter-layer weak. In an essentially 2-dimensional (2-D) system such as thin
films, only the in-plane interactions are important and the control of the
orbital ordering should thus allow the control of magnetic exchange and the
associated Curie temperature, $T_c$. According to the above argument, the
Curie temperature of 2-D thin films will be maximized if one (i) stabilizes
the $d_{x^2-y^2}$ orbital over the $d_{3z^2-r^2}$ and (ii) maximizes the
in-plane effective exchange integral, between the $d_{x^2-y^2}$ orbitals, by
preventing the tilt of the octahedra observed in the
bulk~\cite{structLSMO}. Indeed, in the double exchange mechanism, $T_c$ scales
as
\begin{equation}
T_c \sim J \sim t_{dp}^2 \sim S_{dp}^2
\end{equation}   
($J$: effective exchange between adjacent Mn ions, $t_{dp}$: $({\rm
  Mn})\,d$--$({\rm O})\, 2p$ transfer integral and $S_{dp}$: overlap) and the
octahedra tilt lowers the $S_{dp}$ overlap between the Mn $d_{x^2-y^2}$ and
the O $2p_\lambda$ bridging orbitals. 
  
$\rm A_{1-x}B_xMnO_3$ thin films deposited on a cubic substrate presenting
$a=b$ parameters slightly larger than the bulk manganite value should thus
insure an increase of the manganite in-plane parameters and, in order to
approximately conserve the cell volume, a decrease of the $c$ parameter. With
$c/a < 1$ the $d_{x^2-y^2}$ orbital occupation and large $T_c$ should be
favored, as desired.

Several authors synthesized such thin films as for instance $\rm
La_{2/3}Sr_{1/3}MnO_3$ (LSMO) thin films deposited on the $\rm SrTiO_3$ cubic
substrate~\cite{FM}. They found
that for a small number of LSMO unit cells in the $\vec c$ direction, the
magnetization as well as the Curie Temperature are strongly decreased compared
to the 370K bulk value~\cite{bulkTcSr}.  From twelve deposited unit cells
however the bulk behavior is approximately recovered with just a slight $T_c$
reduction. It is clear that the expected behavior and thus orbital ordering is
not realized in these films. In fact the quick recovering of the bulk magnetic
behavior tells us that the strains imposed by the substrate are essentially
very quickly relaxed as a function of the film thickness. This is most
probably the case for the octahedra tilt present in the LSMO bulk. One should
thus find a way to maintain the desired strains (in-place cell parameter
elongation and quadratic symmetry) over the whole film. For this purpose one
can think to alternate LSMO layers with layers of an adequate cubic material
as in $\rm \left[(La_{2/3}Sr_{1/3}MnO_3)_n (BaTiO_3)_p \right]_q$
super-lattices. We thus explored this possibility using first principle
calculations.

\begin{figure}[h]
\resizebox{8.5cm}{!}{\includegraphics{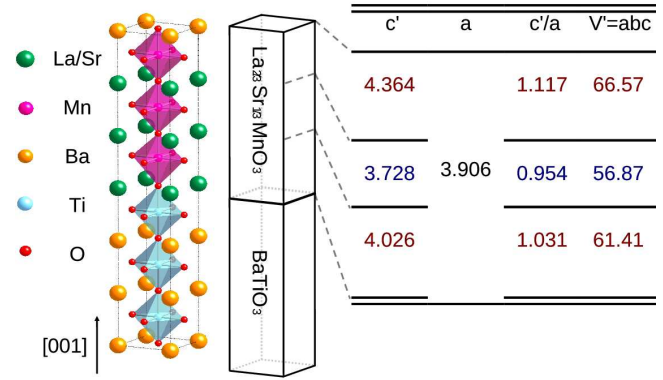}}
\caption{(Color Online) Schematic representation of the super-lattice and
  calculated {
inter-planar distances, c', and associated sub-cell
    volumes, V'} for the LSMO layer (\AA{}).}
\label{fig:SR}
\end{figure}

We studied, using periodic density functional theory~\cite{DFT}, the $\rm
(La_{2/3}Sr_{1/3}MnO_3)_3(BaTiO_3)_3$ super-lattice deposited on a lattice
matched $\rm SrTiO_3$ (001)-oriented substrate (see
figure~\ref{fig:SR}). {
Since the epitaxial films normally follow
  the structure of the perovskite substrate,} we imposed to our system the
substrate in-plane lattice constants, $a=b=3.9056\rm~\AA{}$.  The $c$
parameter was optimized as well as atomic positions.

The super-lattice is built from metallic LSMO layers and insulating $\rm
BaTiO_3$ (BTO) layers. It is thus of crucial importance for the reliability of
the results to be able to accurately position the LSMO Fermi level referring
to the BTO gap. For this purpose we used hybrid functionals known to correctly
reproduce insulating gaps, that is B3LYP~\cite{B3LYP} and the very recent
B1WC~\cite{B1WC} specifically designed to obtain the correct properties on
ferroelectric materials such as $\rm BaTiO_3$. We thus used the CRYSTAL
package~\cite{CRYSTAL} with the basis sets and effective core pseudo-potentials
(ECP) of reference~\onlinecite{Bases}.
Another important technical point is the treatment of the $\rm La/Sr$
disorder. Indeed, in periodic calculations it is not possible to properly
mimic the ionic disorder. We thus made a whole set of calculations associated
with the two extreme cases, that is with the $\rm Sr$ ions totally ordered (3
different positions) and with average $\rm La/Sr$ ions. These average ions
were modeled either using ordered $\rm La/Sr$ ECPs but with averaged
effective nuclear charges (3 possibilities)
or by using only $\rm La$ ECPs with average effective charges.
All 7 different approximations of the ionic disorder were computed with the
two functionals. All of them yielded equivalent results and physical
properties, even if one finds slight numerical differences. One set of
numerical values (corresponding to B3LYP functional, ordered ECPs with the
strontium in the middle layer and averaged charges) are presented as a matter
of example.

Considering the fact that the LSMO layers are extended in the $(\vec a,\vec
b)$ directions compared to the bulk ($a=b=c=3.87~\rm\AA{}$), our results
(figure~\ref{fig:SR}) are at first quite surprising. Indeed, the two LSMO
layers at the interfaces with the $\rm BaTiO_3$ are not only extended in the
$\vec a$ and $\vec b$ directions, but also in the $\vec c$ direction. It
results a strong extension of the {
associated sub-cell volume}
with respectively $\rm 66.57~\AA{}^3$ ---~for the interface with a Ba-O
layer~--- and $\rm 61.44~\AA{}^3$ ---~for the interface with a Ti-O layer~---
to be compared with the bulk pseudo-cubic unit cell volume of $\rm
58.3~\AA{}^3$~\cite{structLSMO}. On the contrary, the central LSMO layer is
strongly contracted in the $\vec c$ direction resulting in a reduction of the
associated sub-cell volume ($\rm 56.87~\AA{}^3$). One sees immediately that
for the interface layers the $c'/a >1$ anisotropy strongly favors the
stabilization of the $d_{3z^2-r^2}$ orbital. This stabilization is indeed
associated in our calculations with a larger occupation of the $d_{3z^2-r^2}$
orbital compared to the $d_{x^2-y^2}$ one. The anisotropy is reversed for the
central layer ($c'/a<1$) and thus the $d_{x^2-y^2}$ orbital is stabilized and
presents a larger occupation number.  The magnetism is essentially
two-dimensional, {
supported by the Mn ions (with 3.467$\mu_B$,
  3.541$\mu_B$, 3.886$\mu_B$)}. The $\rm BaTiO_3$ layers remain essentially
non magnetic.  We can thus predict two magnetic behaviors in these
super-lattices. The first one is related to the interfacial layers and is
associated with a weak in-plane double-exchange and thus a weak Curie's
temperature ($T_{c 1}$). The second one is related to the central layers and
is associated with a strong in-plane double-exchange and thus a strong Curie's
temperature ($T_{c 2}$). One can thus predict three different phases, the
first one at very low temperature with all three LSMO layers participating to
the magnetization, the second one at larger temperature with only the central
layer participating to the magnetism and finally, at high temperature, the
paramagnetic phase.

{
In an attempt to estimate the order of magnitude of $T_{c 1} /
  T_{c 2}$, one can estimate the double-exchange magnetic coupling. Indeed,
  even if in bulk manganites, the double-exchange is not always sufficient to
  explain the physics and polaron formation should be
  considered~\cite{Millis}, in fully-strained films such as the present one,
  we do not expect such relaxation mechanism to take place.  The
  double-exchange integral is directly proportional to the delocalization
  between adjacent magnetic ions and thus to the square of the overlap
  integral between the magnetic $3d$ orbital and the ligand $2p$ bridging
  orbital.  The $T_{c 1} / T_{c 2}\simeq \left| \langle {\rm
      Mn}\,d_{3z^2-r^2}|{\rm O}\,2p_x\rangle / \langle {\rm Mn}\,d_{x^2-y^2}{\rm
      O}\,2p_x\rangle \right|^2$ ratio can thus be undervalued by
  $1/(4\sqrt{3})^2 = 1/48$. Of course, this is a  calculation using crude
  approximations, however it tells us that there is more than an
  order of magnitude between $T_{c 1}$ and $T_{c 2}$.}

Another important result is the fact that the $\rm BaTiO_3$ triple-layer does
not preserve its insulating character but exhibit a weak, however non null,
density of states at the Fermi level in the dominant LSMO spin
orientation. This electronic delocalization through the $\rm BaTiO_3$
triple-layer insures the ferromagnetic coupling between the different LSMO
layers by what we could call a long-range double-exchange mechanism.

As a conclusion, first principles calculations predict for the super-lattice
\begin{itemize} \itemsep -0.3ex
\item two different ferromagnetic  phases, associated with two Curie
  temperatures differing by  nearly two orders of magnitude, 
\item an optimized high Curie temperature  expected to be larger
  than the bulk one (no octahedra tilt).
\end{itemize}

We thus synthesized $\rm (La_{2/3}Sr_{1/3}MnO_3)_3(BaTiO_3)_3$ super-lattices
using laser-MBE deposition~\cite{Salvador} on a [001]-oriented $\rm SrTiO_3$
substrate. {
We also synthetised similar super-lattices using the
  isoelectronic calcium manganite: $\rm La_{2/3}Ca_{1/3}MnO_3$ (LCMO). Indeed,
  one can expect in the LCMO super-lattice similar orbital ordering effects,
  for similar reasons, as predicted in the strontium one.}

\begin{figure}[h]
\resizebox{6cm}{!}{\includegraphics{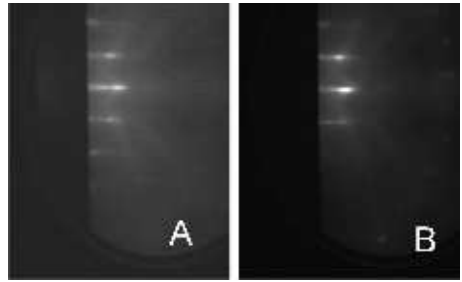}}
\caption{RHEED patterns {
for the BTO/LSMO super-lattice} along the
  [110] azimuth of the $\rm SrTiO_3$ substrate.  A: 620$^\circ$C, $5 \times
  10^{-4}\,$mbar ($\rm O_2 + O_3$). B: 120$^\circ$C, $5 \times
  10^{-7}\,$mbar.}
\label{fig:RHEED}
\end{figure}

The pulsed laser deposition was performed in low pressure ($5.10^{-4}\,$mbar)
to enable control of the growth using the Reflexion High Energy Electron
Diffraction (RHEED). 
While the growth of the structures was carried out in low ozone content (0.1\%
volume), to promote a maximum oxidation of the super-lattice without
favorizing a strong diffusion between the different layers, the structures
were rapidly cooled in high ozone concentration (7\% volume).
%
The deposition of each material, namely BTO and LSMO {
or LCMO},
was measured and calibrated by growing them independently under the same
conditions (620$^\circ$C, pressure $5\% \times 10^{-4}\,$mbar, ozone
concentration 0.1\,\% volume) using an in situ RHEED control for a precise
control of the growth of each layer.
Figure~\ref{fig:RHEED} shows two RHEED patterns, taken along
the [110] azimuth of the substrate. These two pictures evidence a streaky-like
diffraction pattern in agreement with a 2D growth of the super-lattice.
The comparison between the two patterns taken at the end of the deposition (A)
and at the end of the cooling (B) evidences two features.  First, there is no
drastic modification of the pattern after the cooling process. Second the
in-plane structure of the super-lattice exhibits the same lattice parameters
as the $\rm SrTiO_3$ substrate and thus can be described in a perovskite-like
sub-cell. The B pattern also presents Kikuchi lines indicating the high
quality of the deposited film. These lines are not observed in pattern A, due
to the important scattering of the electrons at the deposition pressure.

To achieve a better structural characterization of the super-lattices an X-ray
diffraction study was carried out using a 4-circle diffractometer. Two
different set of reflections were studied: the out of plane reflections and
the so-called asymmetric reflections. From the first set of reflections, the
value of the out-of-plane lattice parameter could be calculated for the $\rm
[(La_{2/3}Sr_{1/3}MnO_3)_3(BaTiO_3)_3]_{25}$ super-lattice.  We measured an
out-of-plane lattice parameter of {
24.281\,\AA{}}, corresponding
almost {
(1\%)} to the value determined from the theoretical
calculations {
(24.519\,\AA{})}.
This value is larger than the
expected value, calculated from the average out-of-plane parameters of films
of the two starting materials.  A Wiliamson-Hall analysis carried out on the
[00l] reflections shows that the coherence length of the diffraction along
that direction corresponds to 600\,\AA{}, that is the thickness of the
deposited film. The same analysis carried out on asymmetrical reflections
([011], [022], ...) indicates that the in-plane lattice parameter is
3.9048\,\AA{}, in agreement with the substrate lattice parameter, and that the
coherence length of the diffraction along an in-plane direction is
1000\,\AA{}, in agreement with the size of the terraces observed on the $\rm
SrTiO_3$ substrates.

\begin{figure}[h]
\resizebox{7cm}{!}{\includegraphics{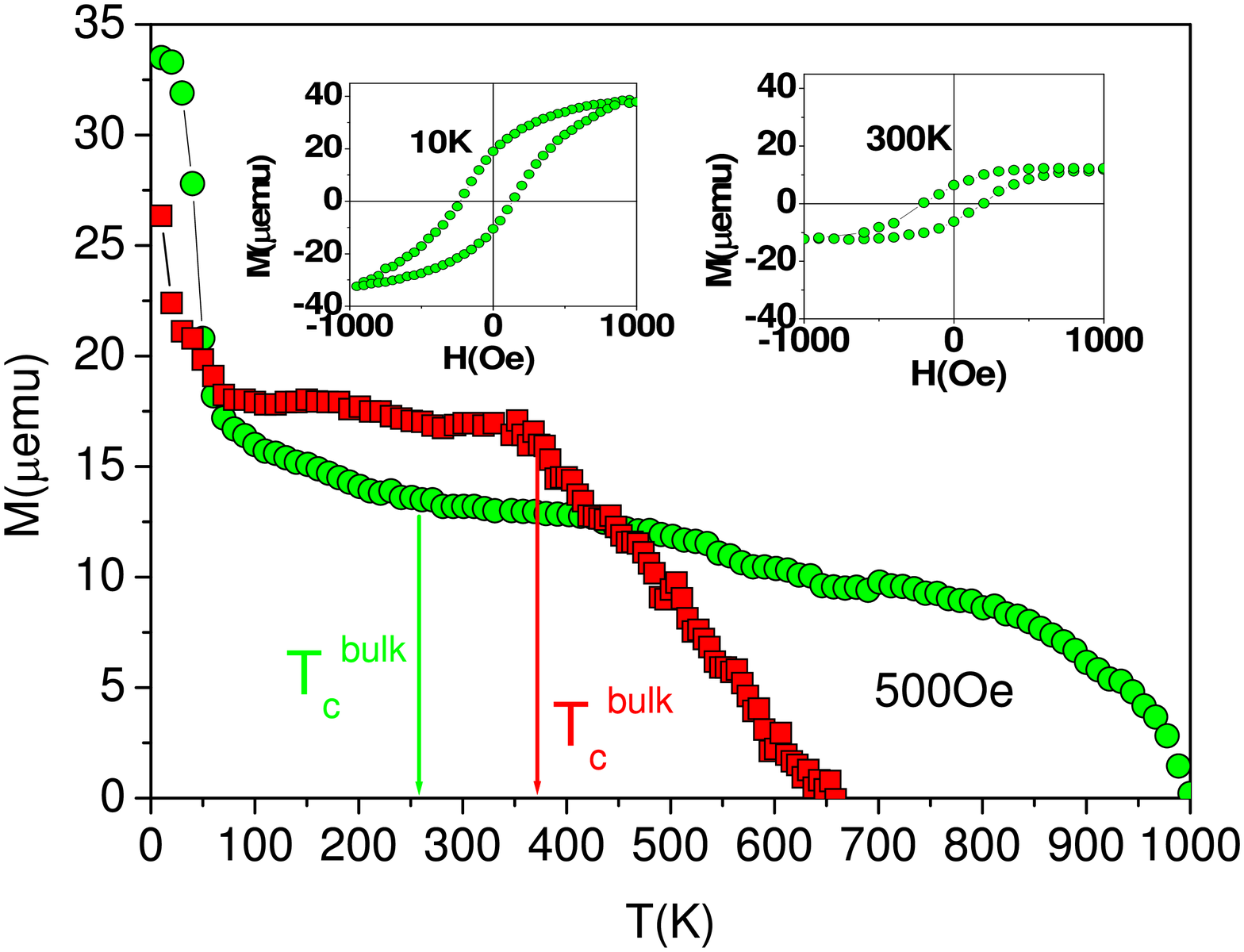}}
\caption{(Color Online) Magnetization (500 Oersteds) for the $\rm
  [(La_{2/3}Sr_{1/3}MnO_3)_3(BaTiO_3)_3]_{25}$ (red squares) and $\rm
  [(La_{2/3}Ca_{1/3}MnO_3)_4(BaTiO_3)_4]_{20}$ (green circles) super-lattices.
  Hysteresis loops of the calcium super-lattice, recorded at 10K and 300K, are
  presented in the inserts. {
The arrows mark the bulk Curie
    temperatures}. The surface of both samples is 16mm$^2$.}
\label{fig:Mag}
\end{figure}

The magnetic measurements (figure~\ref{fig:Mag}), carried out in the
temperature range {
4-1000\,K}, evidence two ferromagnetic phases,
as predicted.  {
The first one at low temperature and the second
  for $T_{c1} < T < T_{c2}$. The low Curie temperature, expected to correspond
  to the two interfacial manganite layers, is about $T_{c1}\simeq 25\,\rm K$
  for the LSMO super-lattice and $T_{c1}\simeq 50\,\rm K$ for the LCMO
  one. The high Curie temperature, expected to correspond to the central
  manganite layer, is about 1000\,K for the LCMO system and about 650\,K for
  the LSMO one, that is far above both $T_{c1}$ and the bulk Curie
  temperatures, as predicted by theoretical calculations. Indeed,} the Curie
temperature of LSMO bulk~\cite{bulkTcSr} or thin films~\cite{fmTcSr} is 370\,K
and in the range 250-260\,K for LCMO bulk~\cite{LSMO-DPh,bulkTcCa} and thin
films~\cite{fmTcCa}. The two inserts of figure~\ref{fig:Mag} show that
magnetic loops can be registered both at 10\,K and 300\,K, {
  insuring true ferromagnetic phases}.

In summary we showed that it is possible to increase the Curie temperature of
a manganite ferromagnetic thin film far above the bulk Curie temperature, by
the control of orbital ordering. The key idea to reach this result is that the
strains applied by the substrate are not sufficient by themselves and that
they must be maintained using the regular intercalation of adequate layers.
Our findings open new opportunities for the design and control of magnetism
in artificial structures and pave the way to novel magneto-electronic
applications, including non-volatile magnetic memory working far above room
temperature.

\section*{Acknowledgements}
We aknowledge financial support from the ``programme ANR blanc'' under the
contract SEMOME, the European comunity and the CNRS (France) under STREP
contract MaCoMuFi (NMP3-CT-2006-033221). Computations were run at the IDRIS 
and CRIHAN computing centers under projets n$^\circ$1842 and n$^\circ$2007013.


\end{document}